\documentclass[11pt,epsf]{revtex4}
\usepackage{graphicx}
\begin{document}
{
\begin{center}
\vspace{10mm}
{
\Large\bf
Neutrino-less Double Beta Decay of $^{48}$Ca studied by CaF$_{2}$(Eu) Scintillators\\
}
\end{center}
S.~Umehara$^1$, 
T.~Kishimoto$^1$, 
I.~Ogawa$^1$, 
R.~Hazama$^2$, 
H.~Miyawaki$^1$, 
S.~Yoshida$^3$, 
K.~Matsuoka$^1$, 
K.~Kishimoto$^1$, 
A.~Katsuki$^1$,  
H.~Sakai$^1$, 
D.~Yokoyama$^1$,  
K.~Mukaida$^1$, 
S.~Tomii$^1$, 
Y.~Tatewaki$^1$, 
T.~Kobayashi$^1$, and
A.~Yanagisawa$^1$\\
$^1$Graduate School of Science, Osaka University, 
Toyonaka, Osaka 560-0043, Japan\\
$^2$Graduate School of Engineering, Hiroshima University, 
Higashi-Hiroshima, Hiroshima 739-8527, Japan\\
$^3$Research Center for Neutrino Science, Tohoku University,
Sendai, Miyagi 980-8578, Japan

\begin{abstract}
We searched for the neutrino-less double beta 
decay(0$\nu\beta\beta$) of $^{48}$Ca
by using CaF$_{2}$(Eu) scintillators.
Analysis of their pulse shapes was effective to reduce backgrounds.
No events are observed in the Q$_{\beta\beta}$-value region
for the data of 3394 kg$\cdot$days.
It gives a lower limit (90\% confidence level) of $T^{0\nu\beta\beta}_{1/2}$
$\textgreater$ 2.7 $\times$ 10$^{22}$ year for the half life of 0$\nu\beta\beta$ of $^{48}$Ca.
Combined with our previous data for 1553 kg$\cdot$days \cite{Ogawa2004},
we obtained 
more stringent limit of $T^{0\nu\beta\beta}_{1/2}$ $\textgreater$ 5.8 $\times$ 10$^{22}$ year.
\end{abstract}
\maketitle

0$\nu\beta\beta$ is acquiring great interest 
after the confirmation of neutrino oscillation 
\cite{Kamiokande,SNO,KamLAND} which demonstrated 
nonzero neutrino mass.
Measurement of 0$\nu\beta\beta$
provides a test for the Majorana nature of neutrinos
and gives an absolute scale of the effective neutrino mass.
Many experiments have been carried out so far
and many projects have been proposed.  
A recent review of 0$\nu\beta\beta$ experiments
is presented elsewhere \cite{AvignoneIII2007}.  

Among double beta decay nuclei,
$^{48}$Ca
has an advantage of the highest Q$_{\beta\beta}$-value (4.27 MeV). 
This large Q$_{\beta\beta}$-value gives a large phase-space 
factor to enhance the 0$\nu\beta\beta$ rate 
and the least contribution from natural background radiations
in the energy region of the Q$_{\beta\beta}$-value.  
Therefore good signal to background ratio is ensured 
in the measurement of 0$\nu\beta\beta$.
However 
not many studies have been carried out 
so far \cite{48Ca1966,48Ca1970,48Ca-Beijing,48Ca-TGV,48Ca-NEMO},
since the natural abundance of $^{48}$Ca is only 0.187 \%.

We carried out the measurements of 0${\nu\beta\beta}$
with CaF$_{2}$(Eu) scintillators.
We previously reported a lower limit (90 \% confidence level)
of 1.4 $\times$ 10$^{22}$ year
for the half life of 0$\nu\beta\beta$ of $^{48}$Ca \cite{Ogawa2004}.
The measurement employed the ELEGANT VI system at Oto Cosmo Observatory.
We observed 0 event in the Q$_{\beta\beta}$-value region
although expected background exceeded 1 event
which limits our experimental sensitivity.
In what follows we describe characteristics 
of our measurement to achieve further background reduction.

The ELEGANT VI system consists of 
three kinds of scintillation detectors.
A CaF$_{2}$(Eu) scintillator (45 mm cube)
works as an active source-detector.
We employed 23 CaF$_{2}$(Eu) scintillators,
which contained 7.6 g of $^{48}$Ca.
CaF$_{2}$(pure) and CsI(Tl) scintillators
work as veto counters for CaF$_{2}$(Eu).
Further details of the detector can be found
in ref. \cite{Ogawa2004}.

A charge sensitive ADC (CSADC) and a flash ADC (FADC)
recorded energy and pulse shape of CaF$_{2}$(Eu) signals, respectively.
The energy is obtained by summing the signals from two photomultiplier tubes (PMTs) 
for each CaF$_{2}$(Eu).
The two signals were used for effective background reduction.

Each CaF$_{2}$(Eu) scintillator was calibrated 
by $\gamma$-rays from $^{137}$Cs
and $\beta$- and $\alpha$-rays from $^{214}$Bi and $^{216}$Po
as internal contaminations.
A standard $\gamma$-ray source of $^{137}$Cs 
was also used.
Linearity has been confirmed 
in an energy range from 662 keV to 3.27 MeV  
and assumed to hold to up to the Q$_{\beta\beta}$-value.
An energy resolution was
measured by a peak width of the $\alpha$-rays from $^{216}$Po.
The peak was clearly observed
at 1.3 MeVee (electron equivalent energy).
The energy resolution was assumed to be inversely 
proportional to square root of deposited energy in an energy region up to 
the Q$_{\beta\beta}$-value.
This dependence has been confirmed 
up to 2.33 MeV by using 0$^{+}\rightarrow$ 0$^{+}$ transition in $^{40}$Ca
for a CaF$_{2}$(Eu) scintillator \cite{Ogawa2004}. 
The extrapolated resolution was evaluated to be 4-6 \% 
in FWHM depending on each CaF$_{2}$(Eu) scintillator.

The ELEGANT VI system is able to strongly suppress backgrounds
by the 4$\pi$ active shield and the large Q$_{\beta\beta}$-value of $^{48}$Ca.
Only a few processes are conceivable as backgrounds \cite{Ogawa2004}.
The main background processes 
are due to pile-up events from Bi and Po nuclei 
which are radioactive 
contaminations in the CaF$_{2}$(Eu) scintillators.

The pile-up events
come from the following sequential decays;\\
(a) $^{212}$Bi (Q$_{\beta}$ = 2.25 MeV) 
\hspace*{1.5mm}\raisebox{-1.2mm}{$_{\beta}$}\hspace*{-6mm}
\raisebox{0.5mm}{ $\longrightarrow$} 
$^{212}$Po (Q$_{\alpha}$ = 8.95 MeV, $T_{1/2}$ = 0.299 $\mu$sec)
\hspace*{1.5mm}\raisebox{-1.2mm}{$_{\alpha}$}\hspace*{-6mm}
\raisebox{0.5mm}{ $\longrightarrow$} $^{208}$Pb (Th-chain),\\
(b) $^{214}$Bi (Q$_{\beta}$ = 3.27 MeV)  
\hspace*{1.5mm}\raisebox{-1.2mm}{$_{\beta}$}\hspace*{-6mm}
\raisebox{0.5mm}{ $\longrightarrow$} 
$^{214}$Po (Q$_{\alpha}$ = 7.83 MeV, $T_{1/2}$ = 164 $\mu$sec)
\hspace*{1.5mm}\raisebox{-1.2mm}{$_{\alpha}$}\hspace*{-6mm}
\raisebox{0.5mm}{ $\longrightarrow$} $^{210}$Pb (U-chain).\\
A typical pulse shape of the pile-up event is shown in FIG. \ref{fg:pulse-dp-pulse}.
In particular, the sequential decay (a) is serious 
because $^{212}$Po has the half life of 0.3 $\mu$s 
which is much shorter than the CSADC gate width of 4 $\mu$s
for the 1$\mu$s decay constant of CaF$_{2}$(Eu) signal.
As a consequence,
the sequential decay (a) frequently becomes a pile-up event in the CSADC gate.
The CSADC gives sum energy of $\beta$- and $\alpha$-rays,
which is occasionally close to the Q$_{\beta\beta}$-value.

The pile-up events can be rejected 
by using pulse shape information.
The 100 MHz FADC recorded the pulse shape in a time window of 10 $\mu$sec,
which is long enough for the CaF$_{2}$(Eu) signal.
In order to reduce data size,
only a sum of 46 signals from 23 CaF$_{2}$(Eu) scintillators
was recorded.
An energy threshold for the FADC was 500 keV.

\begin{figure}
\includegraphics[width=8.0cm]{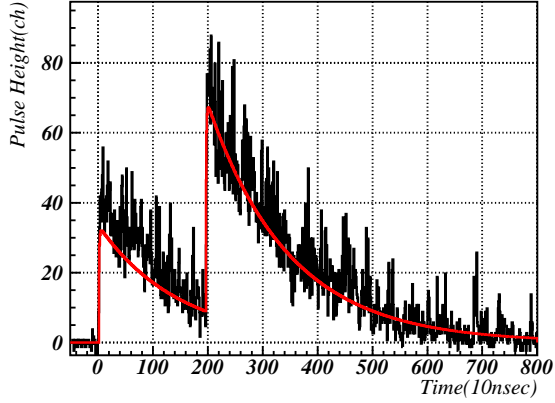}
\vspace*{-5mm}
\caption{\label{fg:pulse-dp-pulse}
A pulse shape of typical  $\beta$-$\alpha$ sequential event 
observed by the FADC.
A solid(thin) line represents pulse shape from the FADC
(fitted pile-up event shape). }
\vspace*{8mm}
\end{figure}

The criteria to select candidate events for 0$\nu\beta\beta$ 
are given as follows.\\
(1) Single CaF$_{2}$(Eu) scintillator fires.\\
(2) No CsI(Tl) scintillators fire.\\
(3) Pulse shape analysis(PSA) tells that the events are not
the pile-up events.\\
The criteria (1) and (2) are the same as those
in the previous analysis \cite{Ogawa2004}.
The new criterion (3) is
to realize further background reduction in this analysis.

PSA can identify the pile-up event
when time difference of the two pulses
is longer than a certain value.
Rejection of the pile-up events is carried out 
by the following three procedures.\\
(i) Preparation of reference pulse shape $f_{ref}(t)$.\\
(ii) Identification of the pile-up events.\\
(iii) Estimation of a rejection efficiency.\\
We describe each step in the followings.

(i) Reference pulse shape ($f_{ref}^i(t)$) was obtained
for each CaF$_{2}$(Eu) scintillator
where $i$ stands for the scintillator number.
Eq.(\ref{eq:expfunction}) represents the pulse shape
where both decay and rise are represented by exponential functions
with time constants $\tau_d^i$ and $\tau_r^i$, respectively.\\
\begin{equation}
\begin{array}{rcll}
f_{ref}^i(t)
&=& A\times ( {\rm exp}(- t/ \tau_d^i)-{\rm exp}(- t/ \tau_{r}^i))&(t \geq 0),\\
&=&0&(t<0).
\end{array}
\label{eq:expfunction}
\end{equation}
Here $A$ is a normalization parameter corresponding to energy.
We fitted the shapes
generated by summing up pulse shapes of events 
in an energy region from 1 MeV to 2 MeV
where we can safely take 
that almost all of the events are due to single-pulse events. 
Obtained $\tau_{d}^i$ ranged from 1230 to 1480 nsec for each scintillator.
After fixing $\tau_{d}^i$,
$\tau_{r}^i$ was obtained by averaging that for each event,
since time jitter between events
deteriorated sharp rise time.
$\tau_r^i$ was obtained as 6 - 8 nsec for each scintillator.

\begin{figure}
\includegraphics[width=16.5cm]{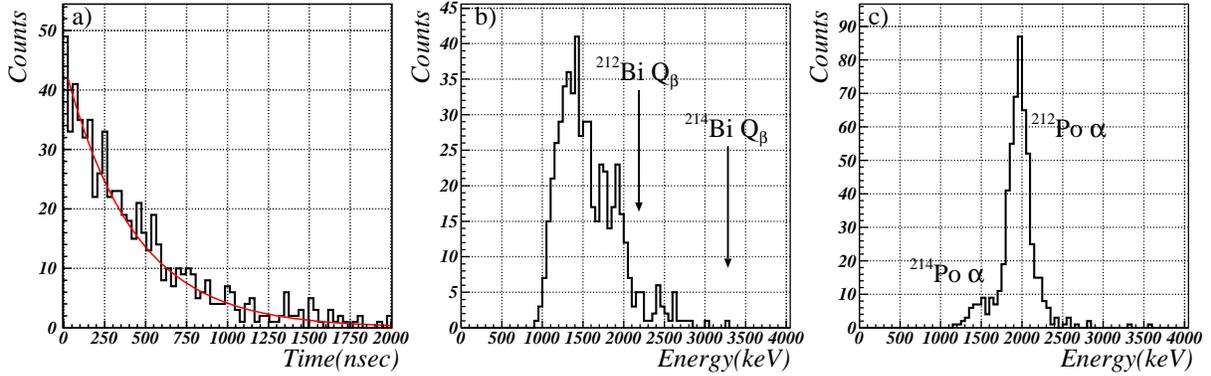}
\vspace*{-10mm}
\caption{
\label{fg:elevi_dp}
Figure a) shows a typical $\Delta t$ distribution
obtained by applying PSA
for events from  3 MeV to 4.5 MeV.
A solid line corresponds to the best fit with 
half life of $T_{1/2}$ = 296$\pm$10 nsec.
Figure b) shows an energy distribution of the prompt component 
for the events
$\Delta t$ $\textgreater$ 30 nsec in figure a).
Q$_{\beta}$-values of $\beta$ decays of $^{212}$Bi and $^{214}$Bi
are indicated by arrows. 
Figure c)
shows an energy distribution of the delayed component for the events
$\Delta t$ $\textgreater$ 30 nsec in figure a).
The electron equivalent energies of $\alpha$-rays
from $^{212}$Po and $^{214}$Po
are 2.0 and 1.6 MeVee, respectively. }
\vspace*{8mm}
\end{figure}

(ii) A fitting function $f(t)$ for the pile-up event
has a delayed component represented by $f_{ref}(t-\Delta t)$

\begin{equation}
\begin{array}{rcll}
 f(t)&= &A_1 \times f_{ref}(t)+ A_2 \times f_{ref}(t-\Delta t)&(t \geq \Delta t),\\
 &= &A_1 \times f_{ref}(t)&(\Delta t \geq t \geq 0),\\
 &=&0&(0>t).
\end{array}
\label{eq:dpfunction}
\end{equation}
Here $A_1$ and $A_2$ correspond to energies of prompt and delayed components,
respectively.
We measured pulse shapes of $\beta$ and $\alpha$ components
and found that their difference
is negligibly small for the present analysis.
We evaluate $\chi ^{2}$ for a certain $\Delta t$ 
by fitting $f(t)$ to events above 3 MeV.
We took $\Delta t$ that gave the least $\chi ^{2}$.

An obtained $\Delta t$ distribution is shown in FIG. \ref{fg:elevi_dp} a).
The $\Delta t$ distribution is well
represented by an exponential decay.
An obtained half life of 296 $\pm$ 10 nsec
is consistent with the half life of 299 nsec of $^{212}$Bi.
The energy spectra of prompt and delayed components
are shown in FIG. \ref{fg:elevi_dp} b) and c).
One can see an end point of 2.2 MeV,
which is consistent with $^{212}$Bi $\beta$ decay
in the spectrum of the prompt component (FIG. \ref{fg:elevi_dp} b)). 
A peak of $\alpha$-rays from $^{212}$Po 
is observed at 2.0 MeVee in FIG. \ref{fg:elevi_dp} c).
These facts show that events above 3 MeV are 
due to the sequential decay (a).

\begin{table}
\caption{\label{tb:0nbb_ana}
A summary of measurements.}
\begin{tabular}{lcc} \hline  
\makebox[40mm]{   } &
\makebox[40mm]{Present measurement} &
\makebox[40mm]{Previous measurement \cite{Ogawa2004}}\\\hline  
Pulse shape information     & yes & no         \\ 
Observed events (counts)    &    0 &     0       \\
Expected background events (counts)&  0.70&   1.30  \\
Statistics (kg$\cdot$day) & 3394 & 1553      \\ 
Half life ($\times$10$^{22}$ year) & 2.7 & 1.4      \\ \hline  
\end{tabular}
\vspace*{8mm}
\end{table}

(iii)
We evaluated the rejection efficiency of the events from the sequential decay (a).
The rejection efficiency depends on where a cut point on $\Delta t$ is set.
We set the cut point at 30 nsec. 
The rejection efficiency was obtained to be 90 \%
by analyzing 
software-generated events from the sequential decay (a).
The rejection may introduce inefficiency for the single-pulse events
due to misidentification.
It is found to be negligible for the present cut point.

A selection of candidate events was made for 3394 kg$\cdot$days of data
as described above.
The energy spectrum is shown in FIG. \ref{fg:elevi_simbg2}.
One finds that the event rate is reduced
by one order-of-magnitude by requiring PSA.
As the result,
we observe no events in a 0$\nu\beta\beta$ window of 4.17 - 4.37 MeV.

Estimation of a background rate
is needed to derive the half life of 0$\nu\beta\beta$.
We know a contribution from the radioactive contaminations
in the scintillators gives dominant background, 
which is estimated by a Monte Carlo simulation \cite{GEANT3}.
Radioactivities in the CaF$_{2}$(Eu) scintillators
were 0.11 mBq/kg and 1.20 mBq/kg on average
for $^{220}$Rn(Th-chain) and $^{214}$Bi(U-chain), respectively.
The radioactivities in each CaF$_{2}$(Eu) 
have been listed in ref. \cite{Ogawa2004}. 
From the simulation using the measured radioactivities 
and the background rejection efficiency,
we estimated the background rate in the 0$\nu\beta\beta$ window
to be 0.70 events/3394 kg$\cdot$day
as given in TABLE \ref{tb:0nbb_ana}.
A contribution from two neutrino double beta decay (2$\nu\beta\beta$) is found 
to be negligible from the life time 
published elsewhere \cite{48Ca-TPC,48Ca-TGV,48Ca-NEMO}.

The detection efficiency
which includes the acceptance efficiency of PSA
was evaluated by a Monte Carlo simulation.
The efficiency was estimated to be 53 \% for the 0$\nu\beta\beta$ window.
It is dominantly determined by the probability
that two electrons from 0$\nu\beta\beta$
are fully contained in a single CaF$_{2}$(Eu).

Here we discuss systematic errors.
They
are mainly from the uncertainties
in the estimation of following three items.
1) Uncertainty on absolute energy calibration and gain stability
may obscure the 0$\nu\beta\beta$ window.
We found it to be less than 1 \%.
2) Uncertainty on PSA efficiencies was estimated to be 3 \%. 
3) Uncertainty on radioactivities in the CaF$_{2}$(Eu) scintillators
may change the estimation of backgrounds.
It was estimated to be 3 \%.
Uncertainty 1) has no effect,
since we still have no event 
even though we have the energy calibration off by 1 \%.  
Uncertainties 2) and 3) are much smaller than statistical error.
We thus do not take them into account in deriving the half life.

Following a procedure in ref. \cite{Feldman1998},
we derive a lower limit at the 90 \% confidence level (C.L.) on the half life 
to be $T^{0\nu\beta\beta}_{1/2}$
$\textgreater$ 2.7 $\times$ 10$^{22}$ year.
The previous measurement employed essentially the same apparatus
except for the FADC and observed no events in the 0$\nu\beta\beta$ window
for 1553 kg$\cdot$days \cite{Ogawa2004}
as given in TABLE \ref{tb:0nbb_ana}.
We combined these results to give more stringent limit.
Considering with 2.0 events of expected backgrounds,
a combined lower limit with the 90 \% C.L.
is $T^{0\nu\beta\beta}_{1/2}$
$\textgreater$ 5.8 $\times$ 10$^{22}$ year
which is the most stringent limit of 0$\nu\beta\beta$ of $^{48}$Ca.
The half life leads to an upper limit on  
the effective Majorana neutrino mass
$\langle m_{\nu} \rangle <$ (3.5 - 22) eV (90 \% C.L.),
using the nuclear matrix elements given in refs. \cite{Suhonen1998,NME2008}.
We present an experimental sensitivity
since the number of observed events 
is fewer than that of the expected backgrounds.
The sensitivity with the 90 \% C.L.
is 1.8 $\times$ 10$^{22}$ year for the combined measurement.

We have studied 0$\nu\beta\beta$ of $^{48}$Ca
by using the ELEGANT VI system,
which realized
an effective background reduction
of radiations from outside the system.
The FADC achieved a reduction of the backgrounds inside the system.
The lower limit for the half life of 0$\nu\beta\beta$ of $^{48}$Ca
was obtained as $T^{0\nu\beta\beta}_{1/2}$
$\textgreater$ 2.7 $\times$ 10$^{22}$ year (90 \% C.L.).
The further stringent lower limit
of 5.8 $\times$ 10$^{22}$ year (90 \% C.L.)
was obtained
by combining with the previous measurement.

One has to prepare large amount of source material to sense the mass region
suggested by the oscillation experiments.
Although 
the 4$\pi$ active shield and the pulse shape analysis
are shown to be effective to realize the background-free measurement,
the ELEGANT VI system is not suitable to scale up.
These techniques are applied to 
the detector system CANDLES \cite{NOON03,TAUP03,NOON04,LRT2004,TAUP05,INPC07,TAUP07},
which realizes large scalability of the detector size.

This work was supported by KAKENHI
No. 06452030, 10640269, 1404485, 14204026, 15340074, 17104003, 18340062 and 18740150.
The works of R.H., H.M. and S.Y.
were partly supported by JSPS Research Fellowships for Young Scientists. 

\begin{figure}
\includegraphics[width=7.6cm]{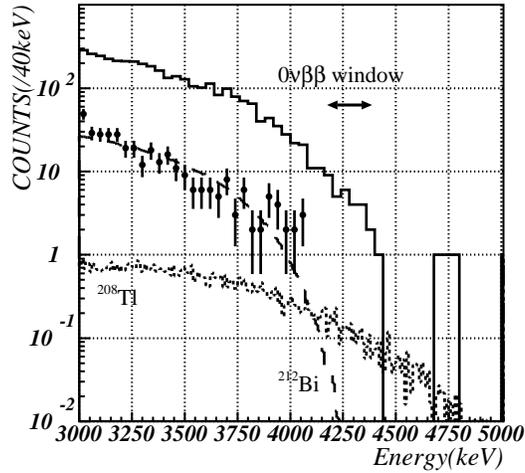}
\caption{\label{fg:elevi_simbg2}
Energy spectra are shown
together with the expected background spectra.
Filled circles represent an experimental data with PSA.
No events are seen in the 0$\nu\beta\beta$ window.
A solid line
represents a spectrum without PSA.
A dashed line and a dotted line
correspond to the expected backgrounds after PSA
from $^{212}$Bi and $^{208}$Tl,
respectively.
Backgrounds from $^{214}$Bi are negligible in this measurement,
because the pile-up events from $^{214}$Bi
were effectively rejected by PSA.}
\vspace*{8mm}
\end{figure}

\bibliography{preprint_0nbb}

\end{document}